\documentclass[conference]{IEEEtran}
\ifCLASSINFOpdf
\else
\fi
\usepackage{graphicx}
\usepackage{tabularx,booktabs}
\usepackage{dingbat}
\usepackage{diagbox}
\usepackage{multirow} 
\usepackage[hyphens]{url}
\usepackage[hidelinks]{hyperref}
\usepackage{xcolor}
\usepackage{amsfonts, amsmath, amsthm, amssymb}
\usepackage{tikz}
\hypersetup{breaklinks=true}
\usepackage{graphicx}
\graphicspath{{figures/}}
\IEEEoverridecommandlockouts
\newcommand\copyrighttext{%
  \footnotesize \textcopyright 2022 IEEE.  Personal use of this material is permitted.  Permission from IEEE must be obtained for all other uses, in any current or future media, including reprinting/republishing this material for advertising or promotional purposes, creating new collective works, for resale or redistribution to servers or lists, or reuse of any copyrighted component of this work in other works.}
\newcommand\copyrightnotice{%
\begin{tikzpicture}[remember picture,overlay]
\node[anchor=south,yshift=10pt] at (current page.south) {\fbox{\parbox{\dimexpr\textwidth-\fboxsep-\fboxrule\relax}{\copyrighttext}}};
\end{tikzpicture}%
}
\begin{document}
\title{Performance Evaluation of Low-Latency Live Streaming of MPEG-DASH UHD video over Commercial 5G NSA/SA Network}

\author{
    \IEEEauthorblockN{Kasidis ARUNRUANGSIRILERT\IEEEauthorrefmark{1}, Bo WEI\IEEEauthorrefmark{1}, Hang SONG\IEEEauthorrefmark{2}, Jiro KATTO\IEEEauthorrefmark{1}}
    \IEEEauthorblockA{\IEEEauthorrefmark{1}Department of Computer Science and Communications Engineering, Waseda University, Tokyo, Japan}
    \IEEEauthorblockA{\IEEEauthorrefmark{2}School of Engineering, The University of Tokyo, Tokyo, Japan
    \\\{kasidis, weibo, katto\}@katto.comm.waseda.ac.jp, songhang@g.ecc.u-tokyo.ac.jp}
}


%Telephone: (800) 555--1212, Fax: (888) 555--1212}


\maketitle
\copyrightnotice
\begin{abstract}
5G Standalone (SA) is the goal of the 5G evolution, which aims to provide higher throughput and lower latency than the existing LTE network. One of the main applications of 5G is the real-time distribution of Ultra High-Definition (UHD) content with a resolution of 4K or 8K. In Q2/2021, Advanced Info Service (AIS), the biggest operator in Thailand, launched 5G SA, providing both 5G SA/NSA service nationwide in addition to the existing LTE network. While many parts of the world are still in process of rolling out the first phase of 5G in Non-Standalone (NSA) mode, 5G SA in Thailand already covers more than 76\% of the population. 

In this paper, UHD video will be a real-time live streaming via MPEG-DASH over different mobile network technologies with minimal buffer size to provide the lowest latency. Then, performance such as the number of dropped segments, MAC throughput, and latency are evaluated in various situations such as stationary, moving in the urban area, moving at high speed, and also an ideal condition with maximum SINR. It has been found that 5G SA can deliver more than 95\% of the UHD video segment successfully within the required time window in all situations, while 5G NSA produced mixed results depending on the condition of the LTE network. The result also reveals that the LTE network failed to deliver more than 20\% of the video segment within the deadline, which shows that 5G SA is absolutely necessary for low-latency UHD video streaming and 5G NSA may not be good enough for such task as it relies on the legacy control signal.\\
\end{abstract}

\begin{IEEEkeywords}
5G Standalone, 5G SA, low-latency, loss, UHD, live streaming, MPEG-DASH
\end{IEEEkeywords}


% \fi
\IEEEpeerreviewmaketitle

\section{Introduction}
The advancement in technology drives the demand for better radio access networks (RAN). While the existing 4G LTE network may be sufficient for the majority of use cases, the rise of UHD video, VR, AR, Industry Automation, Self Driving Vehicle, Internet of Things, etc, requires significantly higher throughput and much lower latency beyond the capability of the existing network \cite{dahmen-lhuissier}. To cope with these issues and lay a strong foundation for the implementation of new technologies, the mobile network has been evolved to the fifth generation (5G), which promises up to 20 Gbps peak downlink throughput and 10 Gbps peak uplink throughput with latency as low as 1 ms for ultra-reliable low-latency communication (URLLC) and also high mobility is a part of the requirements \cite{itu_m2083}\cite{8886119}\cite{9045247}.

\begin{figure}[h!]
  \centering
  \includegraphics[width=0.45\textwidth]{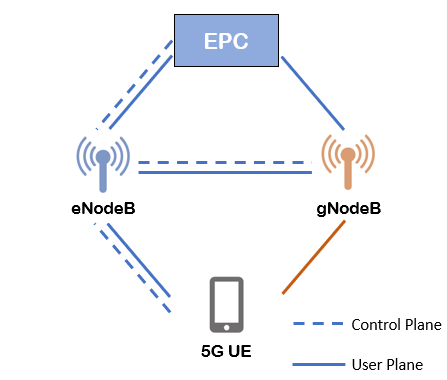}
  \caption{5G Non-Standalone (NSA) Topology}
  \label{fig:5G_NSA_Diagram}
\end{figure}

While the ultimate goal of 5G evolution is to migrate everything, from radio access to the core to the newer generation. The cost and complexity to roll out such big changes in a short period of time are proved to be impossible. With the demand for the new generation network growing rapidly, 3GPP proposed two 5G deployment types in its Release 15 \cite{3GPP_38-912}:
\begin{itemize}
  \item 5G Non-Standalone (NSA): By utilizing LTE Evolved Packet Core (EPC) and LTE RAN as the anchor and provide mobility, coverage, and signaling, then adding 5G as the Secondary Carrier Group (SCG), the deployment of 5G Core (5GC) can be avoided in the initial stage of 5G deployment as shown in Figure \ref{fig:5G_NSA_Diagram}. Therefore, operators can roll out 5G in a shorter period of time with reduced cost.
  \item 5G Standalone (SA): 5GC is being deployed to permit full-scale 5G deployment with new features and functionalities that 5G promise such as Network Slicing, URLLC, and Multi-Gbps support \cite{9069723}.
\end{itemize}
\begin{figure}[h!]
  \centering
  \includegraphics[width=0.45\textwidth]{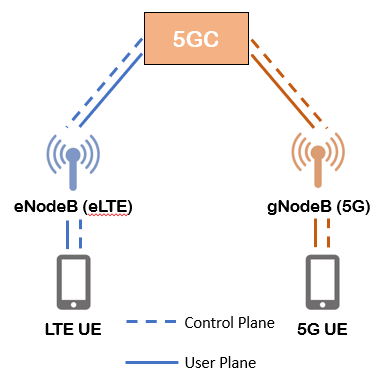}
  \caption{5G Standalone (SA) Topology}
  \label{fig:5G_SA_Diagram}
\end{figure}
In Thailand, the mobile network traffic increased dramatically every year, which drives the demand for 5G networks. The spectrum auction was organized in February 2020 \cite{waring_2020} followed by the nationwide rollout of 5G NSA service on 2500 MHz n41 band by Thailand's biggest operator, AIS, within hours after paying for the spectrum \cite{gsma_2020}. It then quickly followed by the roll-out of low frequency 5G service on 700 MHz n28 band to provide coverage in rural areas. Later on, 5G SA service was operational in parallel with 5G NSA service in July 2021 \cite{sbeglia_2021}. Finally, as of December 2021, more than 76\% of the Thai population are covered by 5G \cite{mgr_online_2021}. Therefore, AIS has 5G SA, 5G NSA, and LTE networks with reasonable coverage, allowing the evaluation and comparison between each network technology.

While there are many solutions for real-time UHD video streaming, one of the most popular standards that allow for adaptive video streaming is MPEG-DASH \cite{6077864}\cite{9500784}. Since it is expected that many video streaming platforms will use MPEG-DASH for real-time UHD video streaming widely in the future, and due to the current availability of 5G SA network with good coverage across the globe, there is almost no study about the behavior of MPEG-DASH when used for real-time UHD video streaming over commercial 5G SA network. Therefore, in this paper, the UHD video will be live-streamed over MPEG-DASH protocol over three types of network: LTE, 5G NSA, and 5G SA, then the performance will be evaluated, compared, and presented. The rest of this paper is organized as follows. Section II will provide some of the previous works. Section III describes how the experiment is set up and how the performance is evaluated in each situation. Section IV presents experimental results with some analysis. Finally, the conclusion and future work should be discussed in Section V.

\section{Related Work}
Although there is previous work that evaluates 5G throughput in Thailand at the major metro station in Bangkok \cite{9497810}, the study only measures average throughput, latency, and loss when UE is stationary using the synthetic test, which may not represent the behavior of real-time MPEG-DASH video streaming. Additionally, the experiment was conducted before 5G SA was launched commercially in Thailand, which implied that the test was done on 5G NSA, which may not necessarily be the final form of the 5G network.

As for the evaluation of MPEG-DASH behavior over mobile networks, there is a previous study \cite{7396692} on the performance of video streaming using MPEG-DASH, RTSP, and RTMP, in both Video-on-Demand (VoD) and live video streaming cases. But the experiment was conducted over LTE network in ideal lab condition with no interference from neighbor cell and UE was at stationary. The video used was also a basic 720p video with a bitrate of 2,500 kbps, which is significantly lower than the requirement of the UHD video. Therefore, the performance of MPEG-DASH when used for live streaming with UHD payload over commercial 5G SA network shall be investigated.

\section{Experiment Setup}


% I wish you the best of success. Please see Figure~\ref{fig:BlockDiagram}.

\subsection{User Equipment (UE)}
To get the full potential of the network, UE that is compatible with 5G SA, 5G NSA, and LTE network in Thailand is needed. Samsung Galaxy Z Flip3 5G (SM-F711B) was used as the UE for the experiment. This UE has Qualcomm Snapdragon 888 Octa-Core processor with 8GB of RAM and runs Android 11 as the operating system. The UE also contains Qualcomm Snapdragon X60 5G RF Modem, which supports 5G SA, 5G NSA, and also LTE mobile network. Since the streaming will be done using software on the laptop, the UE was connected to a laptop via the USB interface in tethering mode to provide internet connectivity to the laptop at the link speed of 425.9 Mbps. The USB interface also provides modem parameters to the laptop, so parameters such as MAC throughput, signal quality, etc. can be logged using AirScreen, a mobile network drive test software by Qtrun Technologies.

\subsection{Mobile Network Configuration}
Since there is only one mobile network in Thailand that provide reasonable 5G SA coverage, AIS was chosen, and ``5G Serenade Extra Max Speed Package 1199 Baht" plan was used, which allow unlimited access to 5G SA, 5G NSA, and LTE network with the maximum connection speed of 1 Gbps. 

At the time the experiment was conducted, there were two 5G bands in service by AIS: 700 MHz (n28) and 2500 MHz (n41). The 2500 MHz (n41) band was allowed to be deployed since February 2020, while 700 MHz (n28) band deployment started in January 2021 due to deferment by the regulator, NBTC \cite{tortermvasana_2020}, resulting in not very good coverage and poor quality of service. Therefore, for the 5G SA test, the UE was locked to use 5G on 2500 MHz (n41) band only. The network configuration for the 5G SA test can be seen in Table \ref{tab:5G_SA}.
\begin{table}[!htbp]
\caption{Technical Parameter for AIS 5G SA Network}
\centering
\label{tab:5G_SA}
\resizebox{8.7cm}{!}{\begin{tabular}{@{}ll@{}}
\toprule
Parameters                          & Values                                                \\ \midrule
Frequency Band                      & TDD 2500 MHz (n41)                                    \\
NR-ARFCN                            & 511950                                                \\
Bandwidth                           & 60/80 MHz depend on area (100 MHz for AIS D.C.)       \\
TDD Slots Configuration (DL/UL)     & 7/2                                                   \\
TDD Symbols Configuration (DL/UL)   & 6/4 (9/4 for AIS D.C.)                                \\
Subcarrier Spacing (SCS)            & 30 kHz                                                \\
Number of Resource Blocks (RB)      & 162 (60 MHz), 217 (80 MHz), 273 (100 MHz)             \\
Number of Component Carrier (CC)    & 1                                                     \\
Number of Layer                     & 4                                                     \\
gNodeB MIMO                         & 64T64R Massive MIMO Split into 8 Beams of 8T8R        \\ 
                                    & (4T4R for AIS D.C.)               \\
UE MIMO (gNodeB Tx/Device Rx)       & 4×4                                                   \\
MCS Table (DL/UL)                   & 256QAM/256QAM                                         \\
\bottomrule
\end{tabular}}
\end{table}

As for the 5G NSA experiment, an LTE anchor band is needed for signaling according to 5G NSA architecture. LTE 1800 MHz (Band 3) was chosen as the anchor band, which resulted in DC\_3A-n41 EN-DC combination, for this experiment because it is the default anchor band for 5G NSA on AIS, it has the widest bandwidth among available LTE Bands on AIS, and it's the most supported combination among commercially available UE as Chinese network also use this combination. The technical parameter for the 5G NSA network test on AIS can be found in Table \ref{tab:5G_NSA}.
\begin{table}[!htbp]
\caption{Technical Parameter for AIS 5G NSA Network}
\centering
\label{tab:5G_NSA}
\resizebox{8.7cm}{!}{\begin{tabular}{@{}ll@{}}
\toprule
Parameters                          & Values                                                \\ \midrule
4G LTE Parameters \\
Frequency Band                      & FDD 1800 MHz (LTE Band 3)                              \\
EARFCN                              & 1450                                                   \\
Bandwidth                           & 20 MHz                                                 \\
Subcarrier Spacing (SCS)            & 15 kHz                                                 \\
Number of Resource Blocks (RB)      & 100                                                    \\
Number of Layer                     & 2 or 4 depending on area                               \\
eNodeB MIMO                         & 2T2R or 4T4R depending on area                         \\
UE MIMO (eNodeB Tx/Device Rx)       & 2×4 or 4×4 depending on area                           \\
MCS Table (DL/UL)                   & 256QAM/64QAM                                           \\
\midrule
5G NR Parameters\\
Frequency Band                      & TDD 2500 MHz (n41)                                    \\
NR-ARFCN                            & 511950                                                \\
Bandwidth                           & 60/80 MHz depend on area (100 MHz for AIS D.C.)       \\
TDD Slots Configuration (DL/UL)     & 7/2                                                   \\
TDD Symbols Configuration (DL/UL)   & 6/4 (9/4 for AIS D.C.)                                \\
Subcarrier Spacing (SCS)            & 30 kHz                                                \\
Number of Resource Blocks (RB)      & 162 (60 MHz), 217 (80 MHz), 273 (100 MHz)             \\
Number of Component Carrier (CC)    & 1                                                     \\
Number of Layer                     & 4                                                     \\
gNodeB MIMO                         & 64T64R Massive MIMO Split into 8 Beams of 8T8R        \\ 
                                    & (4T4R for AIS D.C.)               \\
UE MIMO (gNodeB Tx/Device Rx)       & 44                                                   \\
MCS Table (DL/UL)                   & 256QAM/256QAM                                         \\
\bottomrule
\end{tabular}}
\end{table}
\begin{table}[!htbp]
\caption{Technical Parameter for AIS 4G LTE Network}
\centering
\label{tab:LTE}
\resizebox{8.7cm}{!}{\begin{tabular}{@{}ll@{}}
\toprule
Parameters                          & Values                                                \\ \midrule
Primary Cell (PCell) \\
Frequency Band                      & FDD 1800 MHz (LTE Band 3)                                    \\
EARFCN                            & 1450                                                \\
Bandwidth                           & 20 MHz       \\
Subcarrier Spacing (SCS)            & 15 kHz                                                \\
Number of Resource Blocks (RB)      & 100             \\
Number of Layer                     & 2 or 4 depending on area                               \\
eNodeB MIMO                         & 2T2R or 4T4R depending on area                         \\
UE MIMO (eNodeB Tx/Device Rx)       & 2×4 or 4×4 depending on area                           \\
MCS Table (DL/UL)                   & 256QAM/64QAM                                           \\
\midrule
Secondary Cell 1 (SCell 1) \\
Frequency Band                      & FDD 2100 MHz (LTE Band 1)                                    \\
EARFCN                            & 350                                                \\
Bandwidth                           & 10 MHz       \\
Subcarrier Spacing (SCS)            & 15 kHz                                                \\
Number of Resource Blocks (RB)      & 50             \\
Number of Layer                     & 2 or 4 depending on area                               \\
eNodeB MIMO                         & 2T2R or 4T4R depending on area                         \\
UE MIMO (eNodeB Tx/Device Rx)       & 2×4 or 4×4 depending on area                           \\
MCS Table (DL/UL)                   & 256QAM/64QAM                                           \\
\midrule
Secondary Cell 2 (SCell 2) \\
Frequency Band                      & FDD 2100 MHz (LTE Band 1)                                    \\
EARFCN                            & 550 (525 for AIS D.C.)                                                \\
Bandwidth                           & 10 MHz (15 MHz for AIS D.C.)       \\
Subcarrier Spacing (SCS)            & 15 kHz                                                \\
Number of Resource Blocks (RB)      & 50 (75 for AIS D.C.)             \\
Number of Layer                     & 2 or 4 depending on area                               \\
eNodeB MIMO                         & 2T2R or 4T4R depending on area                         \\
UE MIMO (eNodeB Tx/Device Rx)       & 2×4 or 4×4 depending on area                           \\
MCS Table (DL/UL)                   & 256QAM/64QAM                                           \\
\bottomrule
\end{tabular}}
\end{table}

As for the LTE experiment, AIS allows aggregation of up to three LTE component carriers with two combinations available, CA\_1A-1A-3A and CA\_1A-3A-8A. Both of the combinations offer a total of 40 MHz aggregated bandwidth (except ``AIS D.C." where all combination offers 45 MHz aggregated bandwidth due to one of LTE 2100 MHz carrier having 15 MHz of bandwidth instead of 10 MHz), but MIMO can be different in each carrier when Carrier Aggregation is active. Since LTE 900 MHz (LTE Band 8) is deployed with 2T2R MIMO only. To keep the results consistent, CA\_1A-1A-3A was manually selected. Table \ref{tab:LTE} shows the technical parameter for the LTE experiment.

\subsection{Video Transmission}
The source file was ``70th NHK Kouhaku Uta Gassen" recorded directly from NHK BS4K via BS satellite without any transcoding in m2ts container. The file contains 10-bit HLG H.265/HEVC 3840x2160p59.94 video stream at 25 Mbps with three AAC audio tracks with 2.0, 5.1, and 22.2 channel audio. The file was remux using FFmpeg to mp4 and only 2.0 channel AAC audio was kept. Wowza Streaming Engine was installed on a desktop computer with 1 Gbps (Both Downlink and Uplink) FTTx connection. Wowza Streaming Engine was configured so that it passes through the input RTMP stream directly to MPEG-DASH Streaming Packetizer without any additional transcoding. Since adaptive multi-bitrate streaming is not the objective of this paper, only one video stream in 4K UHD is available. MPEG-DASH Streaming Packetizer was configured so that ``mpegdashChunkDurationTarget" is set to 4000 ms to minimize live latency. Although, 2000 ms, which is equal to YouTube Ultra Low Latency, was also tested, but found to be unstable for 4K UHD live-streaming (YouTube doesn't offer Ultra-Low Latency option for 1440p and 4K resolution \cite{google}).

As for the source, Open Broadcaster Software (OBS) installed on the same PC that runs Wowza Streaming Engine, was used to push video files as RTMP streams to the server. The source video file was played back using OBS internal ``VLC Video Source" source. The encoding parameter was set as seen in Table \ref{tab:OBS_Setup}.

\begin{table}[!htbp]
\caption{Open Broadcaster Software (OBS) Encoding Configuration}
\centering
\label{tab:OBS_Setup}
\resizebox{8.7cm}{!}{\begin{tabular}{@{}ll@{}}
\toprule
Parameters                          & Values                                                \\ \midrule
Video \\
Codec     & H.264/AVC                                                   \\
Encoder                             & NVIDIA NVENC H.264 (new)                              \\
Resolution                          & 3840x2160                                             \\
Frame Rate                          & 60.00p       \\
Bitrate   & 30,000 kbps                                \\
Rate Control Mode     & Constant Bitrate (CBR)                                                   \\
Keyframe Interval            & 2 seconds                                                \\
Encoding Preset      & Max Quality             \\
H.264 Profile    & High                                                     \\
Look-ahead                     & No                                                     \\
Psycho Visual Tuning           & Yes        \\ 
Max B-frames  & 2               \\
\midrule
Audio \\
Codec       & AAC                                                   \\
Sampling Rate                   & 48 kHz                                         \\
Bitrate                   & 320 kbps                                         \\
\bottomrule
\end{tabular}}
\end{table}

\begin{figure}[h!]
  \centering
  \includegraphics[width=0.46\textwidth]{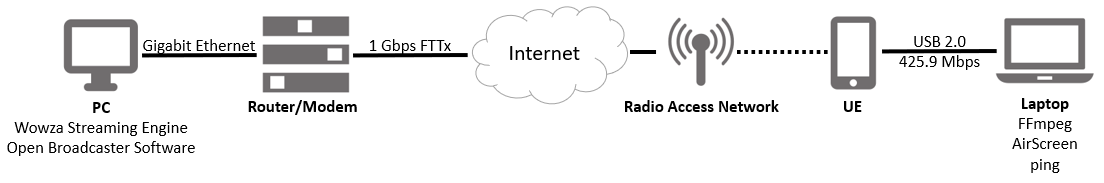}
  \caption{Experiment Topology}
  \label{fig:ExperimentTopology}
\end{figure}

To evaluate the result, FFmpeg was installed on a laptop connected to the UE, then the MPEG-DASH stream was saved to disk in real-time. FFmpeg log was saved to the disk and the number of the skipped segment was counted. Additionally, latency was also measured by using Windows built-in ping command while the stream was being downloaded to the disk. For simplicity, FFmpeg was programmed to automatically terminate after 400 seconds of video has been recorded, which is corresponding to 100 MPEG-DASH segments. Block diagram of the setup can be seen in Figure \ref{fig:ExperimentTopology}.

\subsection{Experiment Scenario}

Six scenarios were evaluated for this paper, four of which is when UE is stationary, while another two is when UE is moving at different average speed. Additionally, baseline result was also obtained by having UE connected to the same network as the MPEG-DASH server via Wi-Fi 6. The access point used was Aruba AP-515 and the channel width was 80 MHz, which provided the link speed of 1,200 Mbps. There was no neighbor access point detected, hence no interference.

\begin{table}[!htbp]
\caption{Summarize of RF condition (Average) in each test case}
\centering
\label{tab:RF_Cases}
\resizebox{8.7cm}{!}{\begin{tabular}{@{}lllll}
\toprule
Test Case  & 5G RSRP & 5G SINR & LTE PCell & LTE PCell \\
 & (dBm) & (dB) & RSRP (dBm) & SINR (dB)\\\midrule
Stationary (Peak) & -101.54 & 16.64 & -101.36 & 2.06\\
Stationary (Off-Peak) & -103.03 & 17.84 & -103.05 & 2.28\\
Stationary (Bad Signal) & -104.52 & 9.30 & -107.63 & 6.83\\
Moving& -88.15 & 16.05 & -82.23 & 8.27\\
Slow Moving & -83.88 & 17.88 & -79.62 & 4.20 \\
AIS D.C. & -78.58 & 44.08 & -72.34 & 21.22\\
\bottomrule
\end{tabular}}

\end{table}
\begin{table*}[!htbp]
\caption{Experiment Result}
\label{tab:Experiment_Result}
\resizebox{\textwidth}{!}{%
\begin{tabular}{@{}lllllllllll}
\toprule
Test Case  & Network & Ping & Ping & Ping & Ping & Ping & Ping & MPEG-DASH & MPEG-DASH & MPEG-DASH\\
& & Avg (ms) & Min (ms) & Max (ms) & Sent & Dropped & Drop \% & Segment Requested & Segment Skipped & Segment Skip \% \\\midrule
Baseline & Wi-Fi 6 & 8 & 4 & 73 & 389 & 0 & 0.00\% & 100 & 0 & 0.00\% \\ \midrule
Stationary (Peak) & 5G SA & 25 & 17 & 81 & 403 & 0 & 0.00\% & 100 & 7 & 7.00\%\\
& 5G NSA & 35 & 24 & 129 & 406 & 0 & 0.00\% & 100 & 27 & 27.00\%\\
& LTE & 39 & 25 & 149 & 401 & 0 & 0.00\% & 100 & 43 & 43.00\%\\\midrule
Stationary (Off-Peak) & 5G SA & 22 & 15 & 52 & 391 & 0 & 0.00\% & 100 & 1 & 1.00\%\\
& 5G NSA & 35 & 24 & 64 & 399 & 0 & 0.00\% & 100 & 5 & 5.00\%\\
& LTE & 32 & 21 & 66 & 402 & 0 & 0.00\% & 100 & 35 & 35.00\%\\\midrule
Stationary (Bad Signal) & 5G SA & 27 & 16 & 128 & 389 & 0 & 0.00\% & 100 & 2 & 2.00\%\\
& 5G NSA & 35 & 23 & 111 & 386 & 0 & 0.00\% & 100 & 14 & 14.00\%\\
& LTE & 36 & 23 & 189 & 406 & 0 & 0.00\% & 100 & 21 & 21.00\%\\\midrule
Moving & 5G SA & 31 & 16 & 217 & 389 & 0 & 0.00\% & 100 & 1 & 1.00\%\\
& 5G NSA & 30 & 17 & 312 & 400 & 0 & 0.00\% & 100 & 6 & 6.00\%\\
& LTE & 62 & 24 & 1875 & 377 & 5 & 1.33\% & 100 & 29 & 29.00\%\\\midrule
Slow Moving & 5G SA & 23 & 15 & 142 & 389 & 0 & 0.00\% & 100 & 1 & 1.00\%\\
& 5G NSA & 30 & 16 & 332 & 407 & 0 & 0.00\% & 100 & 2 & 2.00\%\\
& LTE & 44 & 23 & 550 & 401 & 1 & 0.01\% & 100 & 33 & 33.00\%\\\midrule
AIS D.C. &5G SA & 22 & 17 & 48 & 391 & 0 & 0.00\% & 100 & 3 & 3.00\%\\
& 5G NSA & 26 & 18 & 48 & 415 & 0 & 0.00\% & 100 & 5 & 5.00\%\\
& LTE & 31 & 23 & 56 & 401 & 0 & 0.00\% & 100 & 20 & 20.00\%\\\bottomrule
\end{tabular}%
}
\end{table*}
``Stationary (Peak)" result is obtained with UE place at a Windows on the second floor of a two-story building with medium quality signal, which represents typical use cases. There was no line of sight to the serving eNodeB/gNodeB. Using the timing advance (TA) parameter reveals the distance of approximately 650 m between the UE and the base station. The bandwidth utilized for 5G 2500 MHz (n41) on this serving cell was 60 MHz. The result was obtained during the night peak hour of around 11 PM. ``Stationary (Off-Peak)" was obtained using the same configuration as ``Stationary (Peak)," but was obtained after 1 AM. ``Stationary (Bad Signal)" was obtained at the same location after 2 AM, but UE was moved down to the first floor, which reduce the signal quality significantly.

``Moving" was obtained when UE is moving on Bangna-Trad Expressway at an average speed above 100 km/h. The ``Slow Moving" case was also conducted on an expressway with moderate traffic in downtown Bangkok, which yield an average speed of less than 50 km/h. Since AIS use different 5G bandwidth in a different area, 5G bandwidth is switched between 60 MHz and 80 MHz depending on each cell configuration. In both cases, 51.2\% of the cell has 60 MHz bandwidth, while 48.8\% has a bandwidth of 80 MHz. Lastly, ``AIS D.C." was done inside the AIS Design Centre co-working space, which has a 4T4R distributed antenna system (DAS) and ideal RF condition with a direct line of sight to the antenna. This is also the only place that AIS offers 5G with full 100 MHz bandwidth. Therefore, this test case should be a good representation of the best-case scenario. The summary of RF conditions can be seen in Table \ref{tab:RF_Cases}.

\section{Experiment Results}
To make sure that tethering the internet from the UE to the laptop doesn't affect the ability to stream live UHD MPEG-DASH smoothly, the UE was connected to the same network as the MPEG-DASH server via Wi-Fi, and baseline result was obtained, which shown perfect playback experience with no MPEG-DASH segment skipped with excellent average ping latency. Hence, tethering the internet connection via the USB interface doesn't hinder the test result.

\begin{figure}[h!]
  \centering
  \includegraphics[width=0.46\textwidth]{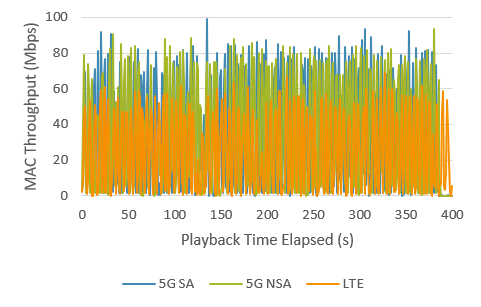}
  \caption{MAC Throughput Characteristic for AIS D.C. Test Case}
  \label{fig:AIS_DC_TP}
\end{figure}

From the experiment, 5G SA showed a significant advantage over 5G NSA in both latency and number of dropped MPEG-DASH segments, demonstrating the latency advantage that the new 5GC and 5G signaling provided. The network was able to respond swiftly and timely to each MPEG-DASH segment request and allocate sufficient results, which provide a smooth playback experience. On the other hand, 5G NSA, which is the widely available type of 5G, has shown a mixed result due to its reliance on LTE signaling. In the case where LTE has a perfect signal quality like the ``AIS D.C." case, 5G NSA may perform only slightly worse than 5G SA with slightly more latency and skipped MPEG-DASH segment as shown in Table \ref{tab:Experiment_Result}. Despite the slight disadvantage, 5G NSA still be able to respond timely for each request of MPEG-DASH segment in this test case and similar throughput characteristic as 5G SA was observed as seen in Figure \ref{fig:AIS_DC_TP}, which explain the experimental result.

\begin{figure}[h!]
  \centering
  \includegraphics[width=0.46\textwidth]{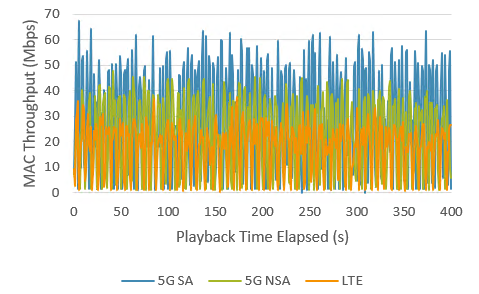}
  \caption{MAC Throughput Characteristic for Stationary (Peak) Test Case}
  \label{fig:Stationary_Peak_TP}
\end{figure}
\begin{figure}[h!]
  \centering
  \includegraphics[width=0.46\textwidth]{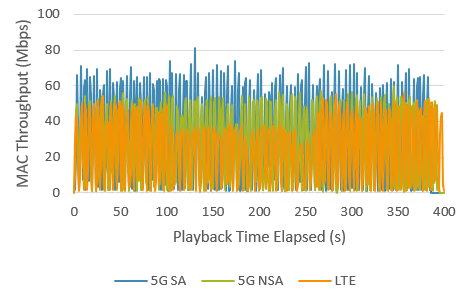}
  \caption{MAC Throughput Characteristic for Stationary (Bad Signal) Test Case}
  \label{fig:Stationary_Bad_TP}
\end{figure}

However, in the cases where LTE is congested or signal quality is poor, 5G NSA performance will be significantly worse than 5G SA as shown in ``Stationary (Peak)" and ``Stationary (Bad Signal)," respectively, where the number of skipped MPEG-DASH segment reach more than 10\%, while 5G SA was able to maintain lower than 3\% skipped MPEG-DASH segment in both cases. The throughput characteristics of the two test cases shown in Figure \ref{fig:Stationary_Peak_TP} and Figure \ref{fig:Stationary_Bad_TP} also demonstrate a significant difference in throughput characteristic between 5G SA and 5G NSA. Especially, in Figure \ref{fig:Stationary_Bad_TP}, where the throughput of 5G NSA and LTE have very similar characteristics, while 5G SA shows a noticeable advantage.

\begin{figure}[h!]
  \centering
  \includegraphics[width=0.46\textwidth]{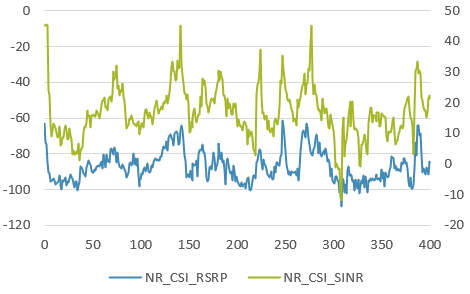}
  \caption{Signal Characteristic for Moving Test Case}
  \label{fig:Moving_Signal}
\end{figure}
\begin{figure}[h!]
  \centering
  \includegraphics[width=0.46\textwidth]{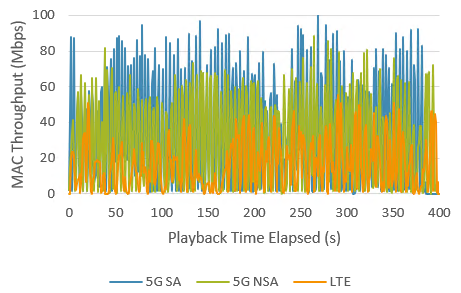}
  \caption{MAC Throughput Characteristic for Moving Test Case}
  \label{fig:Moving_TP}
\end{figure}

The mobility advantage of 5G SA is also demonstrated in the ``Moving" test case, where UE is moving at the average speed of 100 km/h. The signal quality also changes rapidly and wildly as seen in Figure \ref{fig:Moving_Signal}. Despite all of that, 5G SA was able to hold up with only 1\% of skipped segments, respectable ping latency, and also stable throughput as seen in Figure \ref{fig:Moving_TP}. Furthermore, 5G NSA results demonstrate slightly worse performance with as much as six times more skipped MPEG-DASH segment. Lastly, LTE failed the test with almost 30\% of the segment being skipped.


\section{Conclusions and Future Work}

In this paper, the live UHD MPEG-DASH streaming performance of commercial 5G SA, 5G NSA, and LTE network in Thailand was evaluated. Commercially available UE and mobile plans were used for the experiment and the MPEG-DASH Packetizer was tweaked for Ultra Low-Latency live streaming. The results show that 5G SA has both throughput and latency required to deliver a good live UHD streaming experience. The results also show that the LTE network may not be suitable for low-latency live UHD streaming as 20\% or more of MPEG-DASH segments were skipped in all test cases even the ideal one with the perfect signal condition. Interestingly, it has been found that 5G NSA may perform as good as 5G SA in ideal condition or as worst as LTE in poor condition, which shows that the widely available 5G NSA network offered by many operators around the world may not deliver the full potential of 5G and migration to full 5G SA is needed.

Due to financial constraints and complexity, many parts of the world may still stick with 5G NSA deployment for a good while. Therefore, in the future, a workaround for MPEG-DASH segment skipped issue on 5G NSA network shall be purposed and evaluated on commercial 5G NSA network.

\section*{Acknowledgment}

This work was supported in part by the Japan Society for the Promotion of Science KAKENHI (Grant No. 20K14740), in part by Waseda University Grant for Special Research Projects (No. 2021C-132 and No. 2021E-013), in part by NICT (Grant No. 03801), Japan, and in part by JST, PRESTO (Grant No. JPMJPR21PB). Additionally, the authors would like to thank PEI Xiaohong of Qtrun Technologies for providing AirScreen to help with the result collection in this paper.



%\bibliographystyle{IEEEtran}
\Urlmuskip=0mu plus 1mu\relax
\bibliographystyle{IEEEtran}
\bibliography{reference}

\end{document}